\documentclass[10pt]{blois07}
\usepackage{graphicx}
\usepackage{amssymb,amsmath}
\usepackage{cite,./mcite}

\def\mbf#1{\mathchoice{\hbox{\boldmath $\displaystyle #1$}}
        {\hbox{\boldmath $\textstyle #1$}}{\hbox{\boldmath $\scriptstyle #1$}}
        {\hbox{\boldmath $\scriptscriptstyle #1$}}}
\begin{document}
\title{On the Limitations of the Color Dipole Picture}
\author{Carlo Ewerz\,$^a$\protect\footnote{\: speaker at EDS07, Hamburg, May 2007}, 
Andreas von Manteuffel\,$^b$, Otto Nachtmann\,$^b$}
\institute{$^a$ ECT*, Strada delle Tabarelle 286, I-38050 Villazzano (Trento), Italy\\
$^b$ Institut f\"ur Theoretische Physik, Universit\"at Heidelberg, Philosophenweg 16,\\
D-69120 Heidelberg, Germany}
\maketitle

\begin{abstract}
We discuss two aspects of the color dipole picture of high energy 
photon-proton scattering. First we present bounds on various 
ratios of deep inelastic structure functions resulting from 
the dipole picture that, together with the measured data, can be used 
to restrict the kinematical range of its applicability. 
The second issue that we address is the choice of energy variable 
in the dipole-proton cross section. 
\end{abstract}

\section{The dipole picture of high energy photon-proton scattering}
\label{sec:dipole}

The color dipole picture of high energy photon-proton scattering 
(or more generally photon-hadron scattering) 
\cite{Nikolaev:1990ja,Mueller:1993rr} has been a 
very popular and successful framework for the analysis of structure 
function data measured at HERA. In the dipole picture the 
photon-proton scattering is viewed as a two-step process. 
In the first step the real or virtual photon splits into a quark-antiquark 
pair -- a color dipole -- of size $r\!=\!|\mbf{r}|$ in the two-dimensional 
transverse plane of the reaction. The probability for this 
splitting to happen is encoded in the so-called photon wave function 
$\psi^{(q)}_{T,L}(\alpha,\mbf{r},Q)$, where $Q^2$ is the photon 
virtuality, $\alpha$ denotes the fraction of the longitudinal 
momentum of the photon that is carried by the quark, and $q$ indicates 
the quark flavor. In leading order in the electromagnetic and 
strong coupling constants $\alpha_{\rm em}$ and $\alpha_s$ 
the squared photon wave functions for transversely ($T$) and 
longitudinally ($L$) polarized photons are given by 
\begin{equation}
\label{sumpsi+dens}
\left| \psi_T^{(q)} (\alpha, \mbf{r},Q) \right|^2 
= 
\frac{3}{2 \pi^2} \, \alpha_{\rm em} Q_q^2 
\left\{ \left[ \alpha^2 + (1-\alpha)^2 \right] 
\epsilon_q^2 [K_1(\epsilon_q r) ]^2 
+ m_q^2 [K_0(\epsilon_q r) ]^2 
\right\} 
\end{equation}
and 
\begin{equation}
\label{sumpsiLdens}
\left|\psi_L^{(q)}(\alpha, \mbf{r},Q) \right|^2 
=
\frac{6}{\pi^2} \, \alpha_{\rm em} Q_q^2 
Q^2 [\alpha (1-\alpha)]^2 [K_0(\epsilon_q r) ]^2 
\,,
\end{equation}
respectively. Here $m_q$ is the quark mass for flavor $q$, $Q_q$ the 
corresponding electric charge, and 
$\epsilon_q = \sqrt{\alpha (1-\alpha) Q^2 +m_q^2}$. 
$K_{0,1}$ denote modfied Bessel functions, and we have 
summed over the polarizations of the quark and antiquark. 
Integrating over the longitudinal momentum fraction $\alpha$ 
we obtain a density for the photon wave function, 

\begin{equation}
\label{wispsi2}
w^{(q)}_{T,L}(r,Q^2) =
\int^1_0 d\alpha \,
\left|
\psi^{(q)}_{T,L}(\alpha,\mbf{r},Q)
\right|^2 
\,.
\end{equation}
It gives the probability that a highly energetic, transversely or 
longitudinally polarized photon of 
virtuality $Q^2$ splits into a quark-antiquark dipole of flavor $q$ 
and size $r$. 

In the second step of the reaction, the color dipole of size $r$ scatters 
off the proton. Here the dipole is assumed to consist of an on-shell quark 
and antiquark and is treated as a hadron-like state. 
The second step is expressed in terms of the dipole-proton cross 
section $\hat{\sigma}^{(q)}(r,W^2)$ which naturally depends on the 
quark flavor, the size of the dipole, and on the c.m.s.\ energy $W$ of 
this subprocess. The two steps of the photon-proton reaction 
are then connected by integrating over the size and orientation 
of the intermediate dipole state and by summing over quark flavors 
to obtain the total $\gamma^{(*)}p$ cross section, 
 \begin{equation}
\label{sigTLdip}
\sigma_{T,L}(W^2,Q^2)=
\sum_q\int d^2 r \,
w^{(q)}_{T,L}(r,Q^2)\,
\hat{\sigma}^{(q)}(r,W^2) 
\,.
\end{equation}

In general, the dipole cross section $\hat{\sigma}$ 
cannot be calculated from first principles. Instead, one uses 
models for $\hat{\sigma}$ that implement certain 
features like saturation etc., and then fits the parameters 
of these models to measured data for the total structure 
function $F_2$, given for $W^2 \gg Q^2$ and $W^2 \gg m_p^2$ by 
$F_2(W,Q^2) = Q^2 [\sigma_T(W^2,Q^2) 
+ \sigma_L(W^2,Q^2)]/(4 \pi^2 \alpha_{\rm em})$. 

The dipole picture is not exact. Its derivation from a 
genuinely nonperturbative formulation of photon-proton 
scattering -- or, in other words, its foundations in quantum field 
theory -- have been studied in \cite{Ewerz:2004vf,Ewerz:2006vd}. 
As a key result of those papers the assumptions and approximations 
are spelled out in detail which are necessary to arrive at the usual 
dipole picture outlined above. In particular it was possible to 
identify correction terms which are potentially large in certain 
kinematical regions. As with any approximate formula it is 
important to determine as precisely as possible its range 
of applicability -- in the case of the dipole picture the kinematical 
range in which potential corrections to the formulae given above 
are small. We will address this issue in the next two sections. 

Note that the energy variable in the dipole-proton cross section 
$\hat{\sigma}$ is $W^2$. However, many popular models for 
$\hat{\sigma}$ use Bjorken-$x$, $x=Q^2/(W^2+Q^2)$, instead. We will discuss 
the choice of energy variable in section \ref{sec:energyvariable} below. 

\boldmath
\section{Bound on $R=\sigma_L/\sigma_T$}
\unboldmath
\label{sec:boundR}

The densities $w_{T,L}$ are 
obviously non-negative (see (\ref{wispsi2})), and the same holds 
for the dipole cross sections $\hat{\sigma}^{(q)}$, since they are supposed 
to describe the physical scattering process of a dipole on a proton. 
We notice that in the formula (\ref{sigTLdip}) for the cross sections 
$\sigma_L$ and $\sigma_T$ the corresponding densities 
$w^{(q)}_L$ and $w^{(q)}_T$ are convoluted with the same dipole 
cross section $\hat{\sigma}^{(q)}$. Based on these observations 
one can derive bounds on the ratio $R=\sigma_L/\sigma_T$ 
\cite{Ewerz:2006vd,Ewerz:2006an} from the dipole picture. 
The ratio of two integrals 
with non-negative integrands cannot be smaller (larger) than 
the minimum (maximum) of the ratio of the integrands. Applied 
to the cross sections $\sigma_L$ and $\sigma_T$ of (\ref{sigTLdip}) 
this implies 
\begin{equation}
\label{VI55l}
\min_{q,r} \,
\frac{w^{(q)}_L(r,Q^2)}{w^{(q)}_T(r,Q^2)}
\, \leq \, R(W^2,Q^2) 
\, \leq \,\max_{q,r} \,
\frac{w^{(q)}_L(r,Q^2)}{w^{(q)}_T(r,Q^2)} 
\,.
\end{equation}
Note that here the dipole cross sections $\hat{\sigma}^{(q)}$ drop 
out. Consequently, these bounds depend only on the well-known 
wave functions for longitudinally and transversely polarized photons. 
Let us point out that these bounds on $R$ are independent of the choice of 
energy variable ($W^2$ or $x$) in the dipole cross section $\hat{\sigma}$. 
Evaluating the bounds (\ref{sigTLdip}) we find that the lower bound 
is trivial ($R\ge 0$), and the upper bound has the numerical value 
\begin{equation}
\label{boundonR}
R(W^2,Q^2) \le 0.37248
\,.
\end{equation}
This bound has to be satisfied in the kinematical region in which 
the dipole picture is applicable. A violation of the bound in some 
kinematical region, on the other hand, would indicate that the 
dipole picture cannot be used there. 

\begin{figure}
\centerline{\includegraphics[width=9cm]{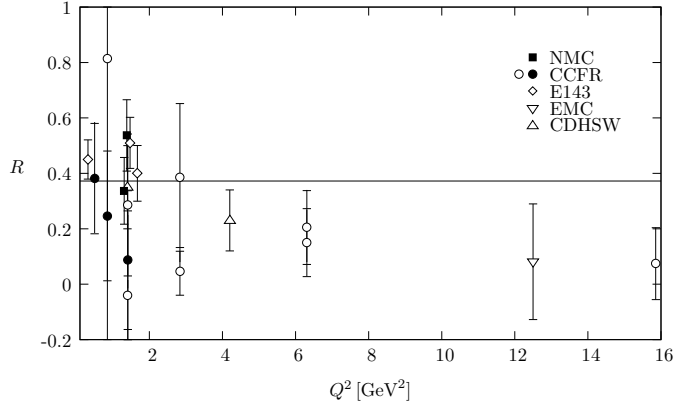}}
\caption{Comparison of experimental data for $R=\sigma_L/\sigma_T$ in the 
region $x <0.05$ with the bound (\ref{boundonR}) resulting from 
the dipole picture. Full points correspond to data with $x <0.01$, 
open points are data with $0.01< x  \le 0.05$. \label{Fig:boundonR}}
\end{figure}

The bound (\ref{boundonR}) is confronted with the experimental 
measurements of $R$ in Fig.\ \ref{Fig:boundonR}, where only data 
points with $x < 0.05$ are included. 
The data have rather large error bars and seem to respect the bound. 
However, in the kinematical region of $Q^2< 2\,\mbox{GeV}^2$ the 
data appear to come very close to the bound -- a situation that 
could hardly be accomodated with a realistic dipole cross section 
$\hat{\sigma}$. The application of the dipole picture in this interesting 
region (in which possible saturation effects are expected to become manifest) 
might therefore be questionable. 
Unfortunately, there are no HERA data on $R$ available which could 
clarify this important point. 
For a detailed discussion and references 
to the corresponding experimental publications see 
\cite{Ewerz:2006vd,Ewerz:2006an}. 

\boldmath
\section{Bounds on ratios of $F_2$ at different $Q^2$}
\unboldmath
\label{sec:boundstructure}

In analogy to the derivation of the bound on $R$ discussed above 
one can also obtain bounds on other ratios of deep inelastic 
structure functions. 
We can for example consider the ratio of structure functions 
$F_2$ taken at the same energy $W$ but at different photon 
virtualities $Q^2$. 
In \cite{Ewerz:2006an,Ewerz:2007md} it was shown that 
for such a ratio one can derive the inequalities 
\begin{equation}
\label{boundonFdiffQ}
\frac{Q_1^2}{Q_2^2} \,\min_{q,r} 
\frac{w_T^{(q)}(r,Q_1^2) + w_L^{(q)}(r,Q_1^2)}
{w_T^{(q)}(r,Q_2^2) + w_L^{(q)}(r,Q_2^2)}
\, \le \,
\frac{F_2(W,Q_1^2)}{F_2(W,Q_2^2)}
\, \le \,
\frac{Q_1^2}{Q_2^2} \,\max_{q,r} 
\frac{w_T^{(q)}(r,Q_1^2) + w_L^{(q)}(r,Q_1^2)}
{w_T^{(q)}(r,Q_2^2) + w_L^{(q)}(r,Q_2^2)}
\,.
\end{equation}
Note that for these bounds to be valid it is essential that the energy 
variable in the dipole cross section $\hat{\sigma}$ is indeed $W^2$, in 
particular, $\hat{\sigma}$ must not depend on $Q^2$. 
These bounds are independent of any other assumptions about 
the dipole cross section $\hat{\sigma}$, and are in fact given 
in terms of the photon wave functions only. They also do not depend 
on the energy $W$. 

\begin{figure}
\centerline{\includegraphics[width=7.7cm]{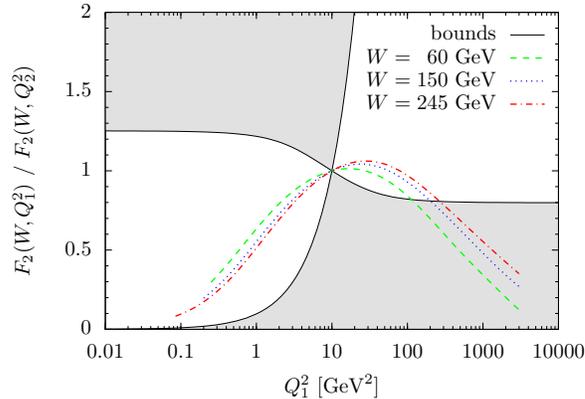}}
\caption{The bounds (\ref{boundonFdiffQ}) on 
$F_2(W,Q_1^2)/F_2(W,Q_2^2)$ for $Q_2^2=10\,\mbox{GeV}^2$ 
and the corresponding fit to HERA data for three different 
values of $W$. Data in the shaded region cannot be described 
in the usual dipole picture. 
\label{Fig:diffQbounds}}
\end{figure}

\begin{figure}
\centerline{\includegraphics[width=7.7cm]{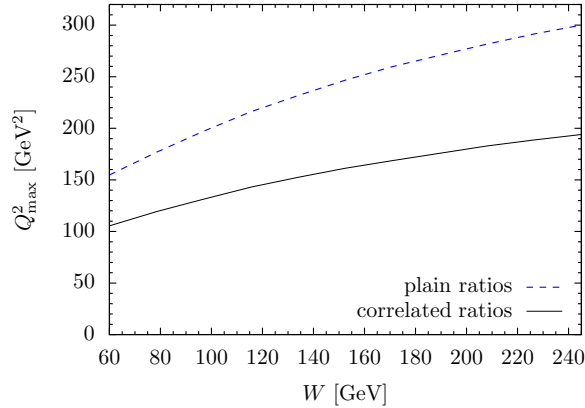}}
\caption{
$W$-dependence of the upper limit $Q^2_{\rm max}$ of the $Q^2$-range 
in which the HERA data are consistent with the bounds obtained from 
the dipole picture. The dashed line results from the bound 
(\ref{boundonFdiffQ}) while the solid line results from correlated bounds 
involving three different values of $Q^2$. 
\label{Fig:2d3dbounds}}
\end{figure}

In this case both the upper and the lower bound are non-trivial. Both 
bounds are shown in Fig.\ \ref{Fig:diffQbounds} for the choice 
$Q_2^2=10\,\mbox{GeV}^2$. 
In the dipole picture the shaded area 
is excluded. Also in that Figure we show the corresponding HERA data 
for three different energies $W$. More precisely, we show the ratios 
resulting from the ALLM97 fit 
\cite{Abramowicz:1991xz,*Abramowicz:1997ms} to the data. 
Within the experimental errors this fit can be regarded 
as a substitute of the data and is more convenient to use for a comparison 
with our bounds. (Note that we use this fit only within the 
kinematical range of the actual data.) As can be seen in the Figure the 
data violate the bound (\ref{boundonFdiffQ}) at large $Q^2$ while the bound 
is respected at low $Q^2$. We can therefore obtain a maximal photon 
virtuality $Q^2_{\rm max}$ beyond which the dipole picture breaks down. 
(In order to obtain an optimal value we have also varied the reference 
scale $Q_2^2$.) 
The $W$-dependence of this maximal $Q^2$ is shown as the dashed 
line in Fig.\ \ref{Fig:2d3dbounds}. As expected, 
the dipole picture can be used up to higher $Q^2$ for larger values of $W$. 

In \cite{Ewerz:2007md} we have considered correlated ratios of 
$F_2$-structure functions taken at the same energy $W$ but at three 
different photon virtualities $Q_i^2$. It turns out that we can derive bounds 
on these correlated ratios from the dipole picture which are stronger 
than the bound discussed above. These bounds can be obtained from 
elementary geometrical considerations, but space limitations prevent 
us from presenting them here. We refer the interested 
reader to \cite{Ewerz:2007md} for a detailed description. Using 
those methods we can show, for instance, that 
$F_2(W,Q_1^2)/F_2(W,Q_3^2)$ is restricted to a certain range that 
in turn depends on the value of $F_2(W,Q_2^2)/F_2(W,Q_3^2)$. 
Also these correlated bounds do not involve any model assumptions 
about the dipole cross section $\hat{\sigma}$ and are entirely 
given in terms of the photon wave functions. 
By confronting the correlated bounds with the ALLM97 fit to HERA 
data we have been able to restrict even further the range in $Q^2$ allowed 
by the dipole picture. More precisely, we have obtained a $Q^2_{\rm max}$ 
up to which the three values $Q_i^2$ can be chosen arbitrarily without 
giving rise to a violation of the bound by the corresponding data. 
In Fig.\ \ref{Fig:2d3dbounds} 
the $W$-dependence of this $Q^2_{\rm max}$ is shown as the solid line. 
The correlated bounds give a stronger restriction on the 
kinematical range in which the dipole picture can be used 
as compared to the bound on the plain ratios (\ref{boundonFdiffQ}). 

\section{The energy variable in the dipole cross section}
\label{sec:energyvariable}

Let us finally turn to the choice of energy variable in the 
dipole cross section $\hat{\sigma}$. Recall that the photon wave function 
describes the probability that a photon of virtuality $Q^2$ splits into 
a dipole of size $r$. Clearly, for a given $Q^2$ dipoles of all possible 
sizes $r$ can emerge, with probabilities given by 
(\ref{sumpsi+dens}), (\ref{sumpsiLdens}) and (\ref{wispsi2}). 
It is therefore not possible to extract $Q^2$ from the dipole 
size $r$. Let us further recall that the second step of the scattering 
process in the dipole picture is the scattering of the dipole of size $r$ 
on the proton. This dipole is fully characterized by $r$ (and -- less 
relevant here -- its longitudinal momentum, $\alpha$ and 
the spin orientations). The dipole-proton cross section $\hat{\sigma}$, 
understood as an actual scattering process of its own, can only depend 
on the properties of the initial state, namely the dipole and the proton. 
In particular, it cannot depend on the photon virtuality $Q^2$. 
Hence $\hat{\sigma}$ cannot be a function of Bjorken-$x$ which can 
only be calculated with the knowledge of $Q^2$. For a more formal 
presentation of this argument see \cite{Ewerz:2006vd}. 

\begin{figure}
\centerline{\includegraphics[width=8cm]{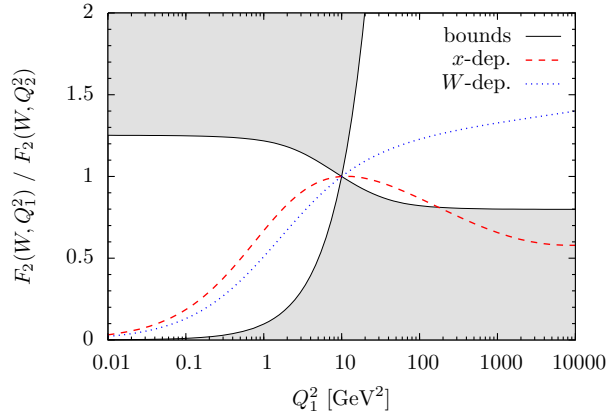}}
\caption{
Ratios of structure functions for the GBW model 
\protect\cite{GolecBiernat:1998js} 
($x$-dependent dipole cross section, dashed line) and for a modification 
of the model with a $W$-dependent dipole cross section (dotted line)
in comparison with the bounds (\ref{boundonFdiffQ}). 
\label{Fig:QRreplace}}
\end{figure}

It is an interesting observation, on the other hand, that in the recent past 
almost all phenomenologically successful models for the dipole cross 
section use $x$ as its energy variable. The most prominent example 
is the Golec-Biernat-W\"usthoff (GBW) model \cite{GolecBiernat:1998js}, 
for further references see \cite{Ewerz:2004vf,Ewerz:2006vd}. 
It is often argued that the probability distribution of dipole sizes 
has a maximum at 
$r \simeq C/Q$ (where $C \simeq 2.4$), and that therefore one can 
effectively replace the dipole size $r$ in $\hat{\sigma}$ 
by its most likely value. The value $Q^2$ in $x$ is then interpreted 
as corresponding to this most likely dipole size $r$, $Q=C/r$. 
In Fig.\ \ref{Fig:QRreplace} we have inverted this procedure 
for the case of the GBW model, that is we have reconstructed a 
$W$- but not $Q^2$-dependent $\hat{\sigma}$ from its $x$-dependence. 
Here we plot again the ratio 
$F_2(W,Q_1^2)/F_2(W,Q_2^2)$, with the choice $W=60\,\mbox{GeV}$ 
and $Q_2^2=10\,\mbox{GeV}^2$, in order to compare 
with the bound (\ref{boundonFdiffQ}). 
The effect of replacing $Q \to C/r$ turns out to be sizable, especially 
at large $Q^2$. The modified model by construction respects 
the bound (\ref{boundonFdiffQ}), while the original GBW model 
strongly violates it at large $Q^2$. The considerable difference 
between the two arises 
because the peak of the distribution of dipole sizes is actually 
rather broad, such that using only its maximum value is 
in fact not a good approximation. 

Strictly speaking, the use of $x$ instead of $W$ in the dipole cross 
section is incorrect 
in the dipole picture. Phrased positively, it is actually a step 
{\sl beyond} the dipole picture to use an $x$-dependent dipole cross 
section $\hat{\sigma}$. The better agreement of the 
$x$-dependent models of the dipole cross section with the data 
seems to indicate that some important corrections to the dipole 
picture are effectively taken into account by using $x$ as the 
energy variable. In our opinion it would 
be very interesting to identify and to understand these additional 
contributions. 

\section{Conclusions}

The dipole picture of high energy photon-proton scattering is only 
an approximation. We have derived various bounds on ratios of 
structure functions from the dipole picture, 
and have used these bounds to restrict the kinematical range of 
applicability of the dipole picture. One should analyse the data in the 
framework of the dipole picture only within this allowed range if 
one wants to arrive at firm conclusions. Further, we have discussed 
the choice of energy variable in models of the dipole cross section. 
We have pointed out that this issue is more delicate than has previously been 
assumed and certainly deserves to be studied in more detail. 

\begin{footnotesize}
\bibliographystyle{blois07} 
{\raggedright
\bibliography{eds07_ewerz}
}
\end{footnotesize}
\end{document}